\documentclass[a4paper, 12pt]{article}
\usepackage[english]{babel}
\usepackage[a4paper, inner=2cm, outer=2cm, top=3cm, bottom=3cm]{geometry}
\usepackage{mathtools}
\usepackage{pstricks}
\usepackage{graphicx}
\usepackage{amsmath}
\usepackage{amssymb}
\usepackage{mathcomp}
\usepackage{textcomp}
\usepackage[doublespacing]{setspace}
\usepackage{etoolbox}
\usepackage{romannum}
\usepackage{caption}

\title{On the effect of renormalization group improvement on the cosmological power spectrum}
\author{R. Moti$^{1}$ and A. Shojai$^{1,2}$\\
$^{1}$\textit{\small Department of Physics, University of Tehran,}\\
$^{2}$\textit{\small Foundations of Physics Group, School of Physics,}\\
\textit{\small Institute for Research in Fundamental Sciences (IPM).}}
\date{}
\begin{document}
\maketitle
\pagenumbering{arabic}
\begin{abstract}
Asymptotically safe quantum gravity predicts running gravitational and cosmological constants, while it remains a meaningful quantum field theory because of the existence of a finite number of non--Gaussian ultraviolet fixed points. Here we have investigated the effect of such running couplings on the cosmological perturbations. We have obtained the improved Mukhanov--Sassaki equation and solved it for two models. The effect of such running of the coupling constants on the cosmological power spectrum is also studied.   
\end{abstract}
\section{Introduction}
The quest to construct a covariant renormalizable quantum gravity attracted more attention in recent years.
Different motivations to quantize gravity is classified by Kiefer \cite{Kiefer} into three categories.
First, the unification. After having a worthy quantum field theory for non--gravitational interactions, unifying quantum theory and general relativity would be a logical wish.
The next motivation comes from singularities of general theory of relativity. It seems that these singularities break down the theory and it is expected to be solved with an appropriate quantum theory of gravity.
Finally the last motivation arises from the fact that there are different concepts of time in quantum theory and general relativity. Time in the latter, is a dynamical object in contrast to quantum theory, in which it is introduced as an external parameter.  Quantum gravity may modify the concept of time in one or both of these theories.

There are various approaches for building quantum theory of gravity.
In \cite{Rovelli}, these approaches are categorized in three sets.
The first category is the \textit{covariant line}, which is simply the quantum field theory of fluctuations over some background metric. This idea leads to higher derivative theories and string theory.
The \textit{canonical line} is the second approach, in which there is no need to introduce the background metric. The Hilbert space can be considered as a representation of operators corresponding to the metric or functions of it. Loop quantum gravity seems to be the latest one in this category.
Finally the name \textit{sum over histories line} is assigned to the third set which contains versions of Feynman's functional integral quantization for the suggested quantum gravity theory. Discrete approaches and spinfoams formalism belong to this set.

In any perturbative approach, which is attractive to particle physicists, offering a finite quantity is an important point in a theory.
One of the difficulties in perturbative quantization of gravity is the negative (mass--) dimension of gravity coupling, i.e. Newton constant, which leads to divergences at short distances or high energies.
Although renormalization of divergences necessarily does not result in a covariant formalism, but would be a vital issue in the suggested quantum gravity theory.
In this context, Weinberg's \textit{asymptotic safety} suggestion in 1979 \cite{1st Weinberg} based on the existence of a finite number of non--Gaussian fixed points, for which the renormalization group (RG) flow attracted by them at infinity, is notable. Many successful research works have been done on the existence of these fixed points (see references in \cite{Weinberg inflation}).
The attraction of the RG--flow to these non--Gaussian UV fixed points at short distances, protects the theory from divergences in that limit.

In 1998, Reuter suggested the truncated exact renormalization group (ERG) method \cite{1st Reuter} for probing non--Gaussian fixed points, where the trajectories of running essential gauge couplings lie on the finite dimensional subspace of the theory space \cite{Finite Theory Space}.
It is known that the $\beta$-functions give us running couplings of the theory which are screened or anti--screened by loop corrections at short distances. 
Treating couplings as fields can be seen in other theories such as scalar--tensor theories (see references in \cite{Varing G}), but here couplings run because of quantization in a systematic manner, i.e. RG equation.
This evolved constant parameter, i.e. running coupling constant, can change the behavior of gravitational phenomena like black holes \cite{IBH}, galaxy rotation \cite{IGR}, CMB and etc.

In this paper, we investigate the effects of the improvement of gravitational and cosmological constants on the Mukhanov--Sassaki equation (MSE).
This equation describes the growth of gauge invariant quantities constructed from quantum perturbations of metric and the inflation scalar field.
These perturbations are usually considered as the primary seeds for inhomogeneities of CMB and the structure formation.
Therefore investigation of the effects of RG improved couplings on the MSE would be remarkable.

The next section is dedicated to a brief introduction of ERG and various improvement methods. Then the improved MSE (IMSE) would be derived in section \ref{Improved perturbations and MSE}. In section \ref{Solutions of IMSE}, we obtain the solution of IMSE for two models, one with a scalar field responsible for the inflation, and one with a cosmological constant. Finally in section \ref{Improved power spectrum}, effects of this improvement on the power spectrum are studied.

It should be noted here that although here we are investigating the improvement of the cosmological power spectrum via the running coupling constants obtained from the asymptotic safe theory, there are other points of view to see the potential impact of renormalization in the power spectrum. For example see \cite{Agullo}.
\section{Truncated ERG in asymptotically safe gravity}
Truncated ERG is one of the several methods for probing non--Gaussian fixed points of gravity theory \cite{Weinberg inflation}.
In this approach, by truncating the scale--dependent effective action $ \Gamma_k[g_{\alpha\beta}] $ up to appropriate interaction terms, other non--effective interaction terms would be ignored.
The evolution of these remaining gauge couplings, which  cannot be eliminated by fields redefinition, is obtained from the exact renormalization group equation (ERGE).
The trajectory of RG flow is
\begin{equation} \label{eq1.1}
    k\partial _{k}\Gamma _{k}=\frac{1}{2}Tr[(\Gamma ^{(2)}_{k}+\mathcal{R}_{k})^{-1}k\partial _{k}\mathcal{R}_{k}]
\end{equation}
where $ \Gamma_k$ integrates out all the fluctuations at scale $ k $ and connects the admissible fundamental action at $ \Gamma _{k \rightarrow \infty} = S $ to the conventional effective action at $ \Gamma _{k \rightarrow 0} = \Gamma $. This average--like effective action at tree--level describes all gravitational phenomena for each momentum of order $ k $ \cite{Bonanno & Reuter1}.
Indeed, the IR--cutoff $ \mathcal{R}_k(p^2) $ which appears in the definition of  $ \Gamma _{k}$, eliminates the effects of fluctuations of $ p^{2} < k^{2} $ on RG flow and is defined by an arbitrary smooth function $ \mathcal{R}_k(p^2) \propto k^2 \mathcal{R}^{(0)}(\frac{p^2}{k^2}) $ where $ \mathcal{R}^{(0)}(\psi) $ satisfies the conditions $ \mathcal{R}^{(0)}(0) =1 $ and $ \mathcal{R}^{(0)}(\psi \rightarrow \infty) \rightarrow 0 $. The exponential form $ \mathcal{R}^{(0)}(\psi) = \frac{\psi}{\exp(\psi) -1} $ is a common chosen form in the literatures \cite{Nagy}. 
Since the multiplicity of couplings in the effective action makes the $\beta$-function intricate, truncation would project the RG flow into the finite dimensional subspace spanned by the essential couplings.
This method gives a finite number of ordinary differential equations.

The Enistein--Hilbert truncation,
\begin{equation}\label{eq1.2}
 \Gamma_k[g_{\alpha\beta}] = \frac{1}{16\pi G_{k}}\int{d^{4}x\sqrt{g}(-R(g)+2\Lambda_{k})}
\end{equation}
is a common truncation for cosmological models.

It is shown numerically in \cite{Bonanno & Reuter2} that for the small values of cutoff $ k \rightarrow 0 $, at \textit{perturbative regime}, the solutions of $\beta$-functions for this model lead to the following power--series dimensionful couplings
\begin{align}
 & G(k) = G_0 \left[ 1 - \omega G_0 k^2 + \mathcal{O}(G_0^2 k^4)\right] \label{eq1.3}\\
 & \Lambda(k) = \Lambda_0 + \nu G_0 k^4 \left[ 1 + \mathcal{O}(G_0 k^2) \right] \label{eq1.4}
\end{align}
where $\omega = \frac{1}{6 \pi} [ 24 \Phi^2_2(0) - \Phi^1_1(0) ]$ , $\nu = \frac{1}{4 \pi} \Phi^1_2(0)$ and $ \Phi^p_n(w) $ is the threshold function, which depends on the IR--cutoff as:
\[
\Phi^p_n(w)=\frac{1}{\Gamma(n)} \int \psi^{n-1} \frac{\mathcal{R}^{(0)}(\psi)-\psi\mathcal{R}^{(0)'}(\psi)}{[\psi+\mathcal{R}^{(0)}(\psi)+w]^p} d\psi \]
$G_{0}$ and $\Lambda_{0}$ are the non--improved coupling constants.

Indeed, at the \textit{fixed point regime}, i.e. $ k \gg m_{pl}$ (where $ m_{pl} = G_0^{-\frac{1}{2}} $ is the Planck mass), the numerical analysis suggests
\[ G(k) = \frac{g_{*}^{UV}}{k^2} \quad , \quad  \Lambda(k) = \lambda_{*}^{UV} k^2  \]
where $ g_{*}^{UV} = 0.27$ and $\lambda_{*}^{UV} = 0.36$ are the non--Gaussian fix point values of dimensionless couplings $ g(k) \equiv k^{2} G(k) $ and $ \lambda(k) \equiv \frac{\Lambda(k)}{k^{2}} $, respectively.

In the following, we are dealing with the end of the inflation epoch with $ k \lesssim  m_{pl}$, and thus it is logical to use the perturbative regime. Hence we use the \eqref{eq1.3} and \eqref{eq1.4} as the running couplings for this period.

As it is the case for any quantum field theory, running couplings are scaled with momentum $k$. On the other hand, in effective field theory, $\Gamma_k$ uses this scaling to specify the cutoff point where the fluctuations with momenta smaller than $k$ are ignored. In the space--time picture this cutoff momentum should be related to the inverse of some physical length. 
Generally speaking, using isotropy and homogeneity of FRW metric, a suitable function $k(t,a(t),\dot{a}(t),...)$ seems to be the best choice for scaling parameter.
Noting that in the standard cosmology, when the Universe is aged $t$, fluctuations of frequency smaller than $1/t$ are ignorable, and also taking into account  the  dimensional considerations, one observes that the cutoff identification
\begin{equation} \label{eq1.5}
  k = \frac{\xi }{t}
\end{equation} 
is a suitable approximation.  Here $\xi$ is a positive constant of order unity.  (This may also be justified for the fixed point regime, where $\Lambda (k) \propto k^2 $., and since we know that $\Lambda \propto H^{2}(t) $. See \cite{Bonanno11}.)
One may also uses the conformal time $\int dt/a(t)$ instead of $t$, but the difference is ignorable because of ignorance of frequencies less than $1/t$. As a result, the above cutoff identification is usually used for improvement of the cosmological models in the literature. For  more details about cutoff identification, the reader is referred to \cite{Bonanno & Reuter1} and \cite{Falls}.

With this identification, we shall have the time dependent parameters $G(k)$ and $ \Lambda(k) $ as:
\begin{align} 
    & G(t) = G_0 \left[ 1- \tilde{\omega}\left(\frac{t_{pl}}{t}\right)^2 + \mathcal{O}\left(\frac{t_{pl}}{t}\right)^4 \right] \label{eq1.6}\\
    & \Lambda(t) = \Lambda_0 + \tilde{\nu} m^2_{pl} \left(\frac{t_{pl}}{t}\right)^4  \left[ 1 + \mathcal{O}\left(\frac{t_{pl}}{t}\right)^2 \right] \label{eq1.7}
\end{align}
where $ \tilde{\omega} \equiv \omega \xi^2 $ and $\tilde{\nu} \equiv \nu \xi^4 $.

The conversion of fundamental units such as  $c$, $ \hbar $ and $G$ to variable ones is a debatable issue \cite{Ellis}. 
In this regard, considering $G$ as a coupling constant, the decision of where and how we should apply the improvement of $G_0$ to $G(x)$ is an important question.

The first and simplest way to do this can be called the \textit{solution improvement}, in which the parameter $G_0$ is replaced by $G(x)$ in any solution of the non--improved theory.
The second approach is \textit{equation improvement}. This is done  at the level of the equations of motion, not the solutions.
The difference between these two methods becomes bold for non--vacuum solutions and the latter seems to be more acceptable if the quantum corrections are negligible in the action.
Generally speaking the improvement of the equations of motion, may leads to solutions different from the former method.

By the third approach which we call the \textit{action parameters improvement} \cite{Reuter & Weyer}, one means substitution of $G_0$ with $G(x)$ in the action, without adding any kinetic term for it.
The improved field equations are obtained from this new action and the externally prescribed field $G(x)$ equation comes from the RGE.
If one adds some kinetic term for it, there would be no guarantee that the obtained $G(x)$ coincides with the result of RGE. Finding a suitable kinetic term is very intricate. 

Here we shall improve the equations of motion, which seems more suitable.
\section{Improved perturbations and MSE}\label{Improved perturbations and MSE}
Metric perturbation during the inflation era and its relation to the matter inhomogeneity is an apt quantum mechanical mechanism  generating initial seeds of structure formation (for a complete review see Ref. \cite{Weinberg-Cosmology}).
The perturbed metric has ten degrees of freedom, but only six of them are physical.
The others, are unphysical because of gauge (coordinate) transformations.
Therefore, introducing gauge invariant quantities, helps us to have results that are independent of the chosen coordinate system.
Among different gauge invariant quantities, the comoving curvature perturbation $R$ would be an appropriate quantity to analyze the power spectrum of the CMB. 
Symmetries of perturbation equations under translations make it easier to work with Fourier modes, $R_q$.

The evolution of $ R_q$ in the non--improved theory follows the MSE \cite{Weinberg-Cosmology}:
\begin{equation} \label{eq2.1}
  \frac{d^2 R_q}{d\tau^2} + \frac{2}{z}\frac{dz}{d\tau}\frac{dR_q}{d\tau}+q^2R_q=0
\end{equation}
where $ \tau = \int^t_{t_0} \frac{dt'}{a(t')} $ is the conformal time and  $z$ which is related to the time derivative of the scalar field as $z=a\dot\varphi/H$ can be considered as the  redshift.
Since the Fourier transformation of 2-point function of the inflation field, i.e. the power spectrum, needs the solution of this differential equation, introducing the initial conditions is unavoidable.

The wave number $ q $ is proportional to the inverse of the comoving wavelength of the perturbations, and
the conformal time is the comoving horizon, hence the quantity $ -q\tau $ will be the ratio of causal horizon length to the comoving  wavelength of the perturbations.
For $ -q\tau \ll  1 $, the wavelength of the perturbations is larger than the length of causal horizon and the perturbations exit the horizon before the end of inflation.
At the end of inflation, with growing the comoving Hubble radius, $ \frac{1}{aH} $, the wavelength of perturbations becomes smaller than the comoving Hubble radius and thus the inhomogeneities and their effect on CMB gradually  become observable.

Here we are interested in investigating the effects of quantum improvements of $G_{0}$ and $\Lambda_{0}$ on these observables. Therefore, at the first step, we have to study briefly the effect of improvement on the perturbation equations. 
Then IMSE in the $ \Lambda$CDM universe will be obtained.
\subsection{Perturbation equations with $ G(t) $ \& $\Lambda(t)$}
The perturbed metric can be written as the sum of the background metric $\bar{g}_{\alpha\beta}$, and the perturbation $h_{\alpha\beta}$, $g_{\alpha\beta}=\bar{g}_{\alpha\beta}+h_{\alpha\beta}$.
We choose the background to be flat homogeneous and isotropic FLRW metric: 
\begin{equation}
  \bar{g}_{00} = -1 \quad , \quad \bar{g}_{0i} = \bar{g}_{i0} = 0 \quad , \quad \bar{g}_{ij} = a^2(t) \delta_{ij}
\end{equation}
with scale factor $a(t)$,  and the perturbation decomposed into (spatial) scalar, vector, and tensor modes.
Hence, the line element becomes
\begin{equation}
  ds^{2} = -(1+E)dt^{2} + a(t) \left[ \partial_{i} F + G_{i} \right] dt dx^{i} + a^{2}(t) \left[\delta_{ij} + A \delta_{ij} + 2 B_{,ij} + 2 C_{(i,j)} + D_{ij} \right]dx^{i}dx^{j}
\end{equation}
where $ A, B, E $ and $ F $ are scalars, $ C_{i} $ and $ G_{i} $ are  divergenceless vectors and $ D_{ij} $ is a symmetric divergenceless--traceless tensor.

Analogous to the metric decomposition, the perfect fluid energy--momentum tensor, $ \bar{T}_{\alpha\beta} = \bar{p}\bar{g}_{\alpha\beta}+(\bar{p}+\bar{\rho}) \bar{u}_{\alpha} \bar{u}_{\beta}$ is perturbed as:
\begin{align}
    & \delta T_{00} = - \bar{\rho} h_{00} +\delta \rho \label{eq2.8}\\
    & \delta T_{i0} = \bar{p} h_{i0} - \left( \bar{\rho} + \bar{P} \right) \left( \partial_{i} \delta u + \delta u_{i}^{V} \right) \label{eq2.9}\\
    & \delta T_{ij} = \bar{p} h_{ij} + a(t)^{2} \left[ \delta_{ij} \delta p + \partial_{i} \partial_{j} \pi^{S}+ 2 \partial_{(i,}\pi_{j)}^{V} + \pi_{ij}^{T} \right] \label{eq2.10}
\end{align}
where a bar over any quantity represents its non--perturbed value.
$ \delta p $ and $ \delta\rho $ are pressure and density perturbations. The velocity perturbation is decomposed as $ \delta u_i = \partial_{i} \delta u + \delta u_{i}^{V} $ to a scalar velocity potential $\delta u $ and a divergenceless vector $\delta u_i $. The other parameters in $\delta T_{ij}$, i.e. scalar $ \pi_{i}^{S} $ , divergenceless vector $ \pi_{i}^{V} $ and symmetric divergenceless-traceless tensor $ \pi_{ij}^{T} $, characterize departure from energy-momentum tensor of perfect fluid.

Since the CMB inhomogeneity comes from scalar perturbation, in the following we omit the vector and tensor perturbations and just consider the scalar mode. Separating these terms on both sides of the Einstein perturbed  equation, $  \delta R_{\alpha\beta} = -8\pi G(t) ( \delta T_{\alpha\beta} - \frac{1}{2} \bar{g}_{\alpha\beta} \delta T^{\lambda}_{\lambda} - \frac{1}{2} h_{\alpha\beta} \bar{T}^\lambda_\lambda ) $, yields:
\begin{align}
    & -4\pi G(t) a^{2} \left[ \delta\rho-\delta p - \nabla^{2}\pi^{S} \right]  =
      \begin{multlined}[t] \frac{1}{2} a \dot{a} \dot{E} + \left(2 \dot{a}^{2} + a\ddot{a} \right) E + \frac{1}{2} \nabla^{2} A \\
      -\frac{1}{2} a^{2} \ddot{A} - 3 a \dot{a} \dot{A} -\frac{1}{2} a \dot{a} \nabla^{2} \dot{B} + \dot{a}\nabla^{2} F \end{multlined} \label{eq2.11}\\ 
    & \partial_{j}\partial_{k} \left[16 \pi G(t) a^{2} \pi^{S} + E + A - a^{2}\ddot{B}-3a\dot{a}\dot{B} + 2a\dot{F} + 4\dot{a}F \right]  = 0 \label{eq2.12}\\
    & 8 \pi G(t) a \left( \bar{\rho}+\bar{p} \right) \partial_{j} \delta u  = - \dot{a} \partial_{j} E + a \partial_{j}\dot{A} \label{eq2.13}\\
    & -4 \pi G(t) \left[ \delta\rho + 3 \delta p + \nabla^{2} \pi^{S} \right]  = 
      \begin{multlined}[t] -\frac{1}{2 a^{2}} \nabla^{2} E - \frac{3 \dot{a}}{2 a} \dot{E} - 
       \frac{1}{a} \nabla^{2}\dot{F} - \frac{\dot{a}}{a^2} \nabla^2 F \\
        + \frac{3}{2} \ddot{A} + \frac{3 \dot{a}}{a} \dot{A} - \frac{3 \ddot{a}}{a} E + \frac{1}{2} \nabla^{2} \ddot{B} + \frac{\dot{a}}{a} \nabla^{2}\dot{B} \label{eq2.14}  \end{multlined}
\end{align}
and the perturbed energy-momentum conservation equation gives:
  \begin{align}
   & \partial_j \left[ \delta p + \nabla^{2} \pi^{S} + \partial_{0} \left[ \left( \bar{\rho} + \bar{p} \right) \delta u \right] + \frac{3 \dot{a}}{a} ( \bar{\rho} + \bar{p} ) \delta u + \frac{1}{2} \left( \bar{\rho} + \bar{p} \right) E \right] = 0 \label{eq2.15}\\
   & \begin{multlined}[t] \delta \dot{\rho} + \frac{3 \dot{a}}{a} \left( \delta\rho + \delta p \right)  + \nabla^{2}\left[ - a^{-1} \left( \bar{\rho} + \bar{p} \right) F + a^{-2} \left( \bar{\rho} + \bar{p} \right) \delta u + \frac{\dot{a}}{a} \pi^S \right] \\ 
   + \frac{1}{2} \left( \bar{\rho} + \bar{p} \right) \partial_{0} \left[ 3 A + \nabla^{2} B \right] = 0 . \label{eq2.16}
  \end{multlined}
  \end{align}  
These are constraints on the perturbation evolutions, by which we shall obtain the IMSE in the next subsection.
Note that the perturbation equations differ from the non--improved ones only in changing $G_{0}$ to $G(t)$ as it is expected in the equation of motion improvement method.
\subsection{IMSE}
Assuming that the inhomogenity of  CMB to be produced by the perturbation of the background inflation field $\bar{\varphi}$,
the action functional would be:
\begin{equation}\label{eq2.17}
   S = S_{\text{Gravity}}+\int d^{4}x \sqrt{-g} \left[ -\frac{1}{2} g^{\alpha\beta} \partial_\alpha \bar{\varphi} \partial_\beta \bar{\varphi} - V(\bar{\varphi}) - 2 M_{pl}^{2} \Lambda \right] .
\end{equation}
where $ M_{pl}^{2} = (8 \pi G_{0} )^{-\frac{1}{2}}$ is the reduced Planck mass.

The dynamics of the model is given by the Friedmann equation for the classical background field $\bar{\varphi}$ :
\begin{equation}\label{eq2.18}
  H^{2} = \frac{8 \pi G_{0}}{3} \left( \bar{\rho}_\varphi + \bar{\rho}_\Lambda \right)
\end{equation}
where $ \bar{\rho}_\varphi = \frac{1}{2} \dot{\bar{\varphi}}^{2} + V(\bar{\varphi}) $ and $ \bar{\rho}_\Lambda = \frac{\Lambda(t)}{8 \pi G(t) } $,
with the improved conservation equation (because of the improved Einstein equation $G_{\mu\nu}=-8\pi G(t) T_{\mu\nu}$):
\begin{equation} \label{eq2.19}
  \dot{\bar{\rho}} = -3 H ( \bar{\rho} + \bar{p} ) - \frac{\dot{G(t)}}{G(t)} \bar{\rho}
\end{equation}
where $ \bar{\rho} =  \bar{\rho}_\varphi +  \bar{\rho}_\Lambda $ and $\bar{p} = \bar{p}_\varphi + \bar{p}_\Lambda $ are the total density and power, respectively.  Note that the dynamics of the background field is assumed to be not affected by the improvement and the effect of RG improvement on the perturbations would be investigated. 

Eliminating the potential of the scalar field using \eqref{eq2.18} and \eqref{eq2.19}, the time evolution of the scalar non--perturbed field would be given by:
\begin{equation} \label{eq2.20}
  \dot{\bar{\varphi}}^{2} = \frac{1}{4 \pi G(t)} \left( H \frac{\dot{G}(t)}{2 G(t)} - \dot{H} \right) .
\end{equation}

In order to get the evolution equation of the gauge invariant quantity, we have to fix the gauge.
We take the same gauge as in \cite{Weinberg-Cosmology} and consider $ \delta\varphi_{q} = 0 $ and $ B_{q} = 0 $ to find $R_q$.
Hence $\delta u $ vanishes.
Indeed, perturbations of pressure and density in this gauge are $ \delta\rho_{\varphi} = \delta p_{\varphi} = -\frac{1}{2} E \dot{\bar{\varphi}}^{2} $.
Therefore energy-momentum conservation \eqref{eq2.16} leads to:
\begin{equation} \label{eq2.22}
  - \frac{2}{a} H \nabla^{2} F - \ddot{A} + \frac{\dot{A}}{\frac{\dot{G}(t)}{2 G(t)}} \left( - \frac{G(t)}{2} \frac{d}{d t} \left(\frac{\dot{G}(t)}{G(t)^{2}} \right) + G(t) \frac{d}{d t} \left( \frac{\dot{H}}{H G(t)} \right) + 3 \dot{H} -\frac{3}{2} H \frac{\dot{G}(t)}{G(t)} \right) = 0.
\end{equation}
It is worth noting that the improvement given by the solutions of RGE, do not change the equations \eqref{eq2.11} and \eqref{eq2.13}, but the conservation equation of the background field changes, as given by \eqref{eq2.22} since the scalar field exchanges energy with the quantum improved part.
This energy exchange can be seen more explicitly in the \eqref{eq2.19} where the dynamic of scalar field is under impression of the improvement term $\frac{1}{\bar{\dot{\varphi}}} \left( \frac{\dot{\Lambda}(t)}{8 \pi G(t)} - \frac{\Lambda(t) \dot{G(t)}}{8 \pi G(t)^{2}} \right) $.

Substituting $ E = \frac{\dot{A}}{H} $ from \eqref{eq2.13} and using \eqref{eq2.11}, the evolution equation of $ A $ can be written in the form:
\begin{equation} \label{eq2.23}
    \ddot{A} + \dot{A} \left( 3 H - \frac{2 \dot{H}}{H} + \frac{\ddot{H}}{\dot{H}} + \frac{\dot{G}(t)}{2 G(t)} \frac{H}{\dot{H}} \frac{d}{d t} \ln \left(\frac{G(t)^{2} \dot{H}}{\dot{G}(t) H^{3}}\right)  \right) - a^{-2} \nabla^{2} A = 0 .
\end{equation}
As mentioned previously, working with Fourier modes is easier, and thus in what follows we use  Fourier transformed equations.

It is clear that the scalar perturbations of metric would not be invariant under transformation of $ x^{\alpha} \rightarrow x^{\alpha} + \xi ^{\alpha} $. The mentioned transformation changes the perturbation $A_{q}$ to $ \tilde{A}_q = A_{q} + H(t)\xi^{0} $. To have a covariant description, usually in the non--improved theory, the gauge invariant quantity $ R_{q} \equiv -\frac{A_{q}}{2} + H\delta u_{q} $  is defined.
Since in this special gauge, the coefficients of $G(t)$ in the \eqref{eq2.11} and \eqref{eq2.13} vanish, the corrections due to running couplings do not affect the scalar perturbations. The corrections manifest themselves in the conservation equation of the background flow. The Hubble parameter $H(t)$, has a correction term  in $R_{q}$ which leads to a second order perturbation in gauge invariant quantity and can be ignored.
Therefore, \textit{the non--improved form of the gauge invariant quantity saves its invariance after improvement of couplings}.
On the other hand the assumption $ -q\tau \ll  1 $ at the late time of inflation, causes the ignorance of the term $ H\delta u_{q} $  with respect to the scalar metric perturbation, $ \frac{A_{q}}{2} $ in $R_{q}$.

Finally considering the conformal time as an independent variable, $ R_{q} $ evolves as:
\begin{equation} \label{eq2.24}
     R_{q}'' + \left( \frac{d}{d\tau} \ln\left( \frac{a H'}{H^{2}}\right ) + \frac{G'}{2 G} \frac{H}{H'} \frac{d}{d\tau} \ln\left(\frac{G^{2} H'}{G' H^{3}}\right)\right)  R_{q}' + q^{2} R_{q} = 0.  
\end{equation} 
Here the prime denotes derivative with respect to the conformal time.
This is the improved Mukhanov--Sassaki equation (IMSE).

Denoting the coefficient of $ R_{q}^{'} $ by $ S(\tau) $:
\begin{equation}
S(\tau)= \frac{d}{d\tau} \ln \left(\frac{a H'}{H^{2}}\right) +  \frac{G'}{2 G} \frac{H}{H'} \frac{d}{d\tau} \ln\left(\frac{G^{2} H'}{G' H^{3}}\right),
\end{equation}
we see that it can be decomposed as:
\begin{equation} \label{eq2.25}
    S(\tau) = S_0(\tau) + S_{I}(\tau)
\end{equation}
in which $ S_0(\tau) = \frac{d}{d\tau} \ln\left( \frac{a H'}{H^{2}} \right) $, and $ S_{I}(\tau) = \frac{G'}{2 G} \frac{H}{H'} \frac{d}{d\tau} \ln\left(\frac{G^{2} H'}{G' H^{3}}\right) $ is the improved part.
It can be seen that in the non--improved limit, where $ G $ and $ \Lambda $ are constants, MSE  \eqref{eq2.1} is recovered.
\section{Solutions of IMSE} \label{Solutions of IMSE}
In this section the solution of the IMSE for two cases will be obtained.
First, for a model in which $ \Lambda_{0} = 0 $ and it remains zero as time passing.
Second, for the $ \Lambda$-inflation model. We shall see that IMSE can be solved iteratively for both cases.
\subsection{Case I: $\Lambda_0=0$}

The Fridemann equation and conservation equation in this case are
\begin{align}  
    & H^{2} = \frac{8 \pi G(t)}{3} \left( \frac{1}{2} \dot{\varphi}^{2} + V(\varphi) \right) \label{eq3.1}\\
    & \dot{H} = -4 \pi G(t) \dot{\varphi}^{2} + \frac{H \dot{G}(t)}{2 G(t)}. \label{eq3.2}
\end{align}
Considering the well--known exponential potential $ V(\varphi) = g e^{-\lambda\varphi}$ , where $ g $ and $ \lambda $ are arbitrary real constants,
the perturbed  $ H = H_{0} +\tilde\omega H_{1} $ and $ \varphi = \varphi_{0} +\tilde\omega \varphi_{1} $  seem to be appropriate solutions for \eqref{eq3.1} and \eqref{eq3.2} up to first order.
Substituting $G(t)$ from \eqref{eq1.6}, gives $ H_{1} $ and its' time evolution, $ \dot{H_{1}} $, up to $ \mathcal{O}(1)$ :
\begin{align}
    & \tilde\omega H_{1}  = \frac{8 \pi G_{0}}{3} \frac{\epsilon}{\lambda}\tilde\omega \dot{\varphi}_{1} - \frac{\lambda}{6} \frac{3-\epsilon}{\epsilon} \frac{\tilde\omega\varphi_{1}}{t} - \frac{8 \pi G_{0}}{3} \frac{\epsilon}{\lambda^{2}} \frac{\tilde\omega t_{pl}^{2}}{t^{3}} - \frac{3 - \epsilon}{6 \epsilon} \frac{\tilde\omega t_{pl}^{2}}{t^{3}} \label{eq3.3} \\
    & \tilde\omega\dot{H_{1}}  = \frac{16 \pi G_{0}}{\lambda^{2}} \frac{\tilde\omega t^{2}_{pl}}{t^{4}} -\frac{16 \pi G_{0}}{\lambda} \frac{\tilde\omega\dot{\varphi}_{1}}{t} + \frac{1}{\epsilon} \frac{\tilde\omega t^{2}_{pl}}{t^{4}} \label{eq3.4}
\end{align}
where $ \epsilon \equiv -\frac{\dot{H}_{0}}{H_{0}^{2}} $ is a positive dimensionless quantity.   
Suggesting power--law solutions of the form $ H_{1} = \frac{b}{t^{3}} $ and $ \varphi_{1} = \frac{c}{t^{2}} $ and determining the coefficients $ b $ and $ c $ by substitution in \eqref{eq3.1} and \eqref{eq3.2}, one obtains the Hubble parameter and the scalar fields in RG improvement approach up to first order as  
\begin{align}
    & \varphi = \frac{1}{\lambda} \ln\left( \frac{8 \pi G_{0} g \epsilon^{2} t^{2} }{3-\epsilon}\right) + \frac{\tilde\omega t^{2}_{pl}}{\left(16 \pi G_{0} \epsilon\right)^{\frac{1}{2}} \left( \epsilon -1 \right) }\frac{1}{t^{2}} \label{eq3.5} \\
    & H = \frac{1}{\epsilon t} - \frac{2\tilde\omega t^{2}_{pl}}{3\left(\epsilon -1\right)} \frac{1}{t^{3}} . \label{eq3.6}
\end{align}   
The independent variable in the IMSE is the conformal time, $ \tau $.
To determine $ t(\tau) $, the scale factor $ a(t) $ is necessary.
Since $ H = \frac{\dot{a}(t)}{a(t)} $, the scale factor becomes
\begin{equation} \label{eq3.7}
    a(t) = \tilde{a}_* t^{\frac{1}{\epsilon}} e^{-\frac{\gamma}{t^2}} 
\end{equation}
where $\gamma =\frac{\tilde\omega t_{pl}^{2}}{3(1-\epsilon)} $, and $\tilde{a}_*$ is a constant. Then the conformal time becomes 
\begin{equation} \label{eq3.8}
    \tau = \int^{t}_{t_{*}\rightarrow \infty} \frac{dt'}{a(t')} = \frac{1}{\tilde{a}_*} \left( \frac{\epsilon}{\epsilon-1} t^{\frac{\epsilon-1}{\epsilon}} - \frac{\epsilon \gamma}{\epsilon+1} t^{-\frac{\epsilon+1}{\epsilon}} \right) .
\end{equation}    
Inverting this, it can be shown that
\begin{equation} \label{eq3.9}
    t = \left( \frac{\tau}{C} \right)^{\frac{1}{1-\frac{1}{\epsilon}}} + \frac{D}{\frac{1}{\epsilon}-1} \left( \frac{\tau}{C} \right)^{\frac{-1}{1-\frac{1}{\epsilon}}}
\end{equation}   
where $ C = \frac{\epsilon}{\epsilon-1} \frac{1}{\tilde{a_{*}}} $ and $ D = \frac{\tilde\omega t^{2}_{pl}}{3(1+\epsilon)} $.

Finally to obtain the IMSE, we can use the above results to get:
\begin{equation}
S_0(\tau)=-\frac{2}{1-\epsilon}\frac{1}{\tau}
\end{equation}
and
\begin{equation}
S_I(\tau)=-K \tau^{\frac{1-3\epsilon}{\epsilon -1}}
\end{equation}
where $ K = \left(\frac{\epsilon}{(\epsilon-1)\tilde{a_{*}}}\right)^{\frac{2 \epsilon}{\epsilon-1}} \left(\frac{4\tilde\omega  \epsilon (2 \epsilon^{2}-3)}{3 (1-\epsilon)(1-\epsilon^{2})}\right) $.

Therefore the IMSE is given by:
\begin{equation} \label{eq3.10}
   R''_{q} - \frac{2}{1-\epsilon} \frac{1}{\tau} R'_{q} + q^{2} R_{q} =  K \tau^{\frac{1-3\epsilon}{\epsilon -1}} R'_{q}.
\end{equation}    

Since the right hand side of IMSE is just a small correction, we can solve it iteratively.
At zeroth order of iteration, ignoring this term, the solution is Hankel functions $H_{\alpha}^{(1)}(-q\tau)$ and $H_{\alpha}^{(2)}(-q\tau)$.
We choose the sub-horizon initial condition  as $ \exp(-iq\tau)$ at large $-q\tau$.
This initial condition is the result of applying WKB methods suggesting plane wave solution for initial conditions at early times, $\frac{q}{a} \gg H $, as in \cite{Weinberg-Cosmology}.
With the conditions:
\begin{equation*}
 \frac{2\dot{\Lambda}(t)}{\dot{G}(t)}-\frac{2\Lambda(t)}{G(t)} \ll  \frac{q^2}{a^2} \quad, \quad\frac{H\dot{G}(t)}{2 G(t)} \ll  \frac{q^2}{a^2} 
\end{equation*}
for $\Lambda$CDM universe, we have the same initial condition as for the non--improved model:
\begin{align*}
  & \delta\varphi_q(t) \rightarrow \frac{1}{(2\pi)^{\frac{3}{2}} a(t) \sqrt{2 q}} \exp(-iq \tau)
\end{align*}
which at the limit $ a(t)\rightarrow 0 $ causes $R_q$ to approach to $- H\frac{\delta\varphi_q}{\dot{\bar\varphi}}$.

Since at large $-q\tau$, $ H^{(1)}_{\alpha} (-q\tau) $ tends to $ \sqrt{\frac{2}{\pi (-q\tau)}} \exp( i(-q\tau) - i\alpha\frac{\pi}{2} - i\frac{\pi}{4} ) $, the first Hankel function would be an appropriate choice.
After renormalizing the zeroth order solution becomes
\begin{equation} \label{eq3.11}
  R_{q_{(0)}}(\tau) = K'(-\tau)^\alpha H^{(1)}_\alpha (-q\tau)
\end{equation}
where $ K' = \frac{-\lambda \sqrt{\pi}}{4 (2\pi)^{\frac{3}{2}} \epsilon}(\frac{\epsilon}{1-\epsilon})^{\frac{-1}{1-\epsilon}} \exp(\frac{i \pi \alpha}{2}+\frac{i \pi}{4}) $ and $ \alpha = \frac{\epsilon-3}{2(\epsilon-1)} $.

To obtain the solution of IMSE up to the next order of iteration we have to substitute the zeroth order solution in the right hand side of (\ref{eq3.10}) and solve the equation. Defining:
\[\mathcal{F}_{\alpha} = (-q)^{3-2\alpha}\int_0^{-q\tau} d(-q\tau)(-q\tau)^{2\alpha-4} J_\alpha(-q\tau) \Big[ \]
\begin{equation} \label{eq3.14}
\left .-q\tau H^{(1)}_{\alpha-1}(-q\tau) + 2\alpha H^{(1)}_{\alpha}(-q\tau)+q\tau H^{(1)}_{\alpha+1}(-q\tau)\right]  + \mathcal{D}_{1(\alpha)}(q)
\end{equation}
\[ \mathcal{G}_{\alpha} =(-q)^{3-2\alpha}\int_0^{-q\tau} d(-q\tau)(-q\tau)^{2\alpha-4} Y_\alpha(-q\tau) \Big[ \]
\begin{equation}\label{eq3.14aa}
\left .-q\tau H^{(1)}_{\alpha-1}(-q\tau) + 2\alpha H^{(1)}_{\alpha}(-q\tau)+q\tau H^{(1)}_{\alpha+1}(-q\tau)\right] + \mathcal{D}_{2(\alpha)}(q)
\end{equation}
in which $ \mathcal{D}_{1(\alpha)}(q)$ and $\mathcal{D}_{2(\alpha)}(q) $ are integration constants, the solution up to first order becomes:
\begin{equation} \label{eq3.15}
R_{q(1)}(\tau) = K' (-\tau)^{\alpha} \left( H^{(1)}_\alpha(-q\tau) -\frac{K \pi}{4} \left[\mathcal{F}_{\alpha} Y_\alpha(-q \tau)+\mathcal{G}_{\alpha} J_\alpha(-q\tau)\right]\right).
\end{equation}

The perturbations outside the horizon are defined by the limit $ -q \tau \ll 1 $.
At this limit, for  $ \alpha > 0 $, the term $J_\alpha(-q\tau)\rightarrow (\frac{-q\tau}{2})^{\alpha}\frac{1}{\Gamma(\alpha+1)} $ can be ignored with respect to $Y_\alpha(-q \tau) \rightarrow \frac{-\Gamma(\alpha)}{\pi} (\frac{-q\tau}{2})^{-\alpha}$.
On the other hand, $\mathcal{F}_\alpha$ at $ -q \tau \ll 1 $ becomes
\begin{equation}
  \mathcal{F}_\alpha = (-q)^{3-2\alpha} ( \mathcal{A}_{\alpha}(-q\tau)^{4\alpha-3} + \mathcal{B}_{\alpha}(-q\tau)^{4\alpha-1} + \mathcal{C}_{\alpha}(-q\tau)^{2\alpha-1} )+ \mathcal{D}_{1(\alpha)}(q)
\end{equation}
with
\begin{align*}
 \mathcal{A}_{\alpha} & = \frac{1}{4\alpha-3}\left(\frac{2^{1-2\alpha}\alpha}{\Gamma(1+\alpha)^2} - \frac{i2^{1-2\alpha}(\cos((-1+\alpha)\pi)\Gamma(1-\alpha)+\alpha \cos(\alpha\pi)\Gamma(-\alpha))}{\pi\Gamma(1+\alpha)} +\frac{2^{1-2\alpha}}{\Gamma(\alpha)\Gamma(1+\alpha)}\right)\\
  \mathcal{B}_{\alpha} & = \frac{1}{4\alpha-1}\Big( \frac{2^{-2\alpha}\alpha}{(1+\alpha)\Gamma(1+\alpha)^2} + \frac{i 2^{-1-2\alpha}\cos((1+\alpha)\pi)\Gamma(-1-\alpha)}{\pi\Gamma(1+\alpha)}+ \frac{2^{-1-2\alpha}(1+2\alpha)}{\alpha(1+\alpha)\Gamma(\alpha)\Gamma(1+\alpha)} \\ 
  & + \frac{i 2^{-1-2\alpha}(1+2\alpha)\cos((-1+\alpha)\pi)\Gamma(1-\alpha)}{\pi\alpha(1+\alpha)\Gamma(1+\alpha)} + \frac{i2^{-2\alpha}\alpha \cos(\alpha\pi)\Gamma(-\alpha)}{\pi(1+\alpha)\Gamma(1+\alpha)} - \frac{2^{-1-2\alpha}}{\Gamma(1+\alpha)\Gamma(2+\alpha)} \Big) \\
 \mathcal{C}_{\alpha} & =\frac{1}{2\alpha-1}\left( -i\frac{-\alpha\Gamma(-1+\alpha)+\alpha^2\Gamma(-1+\alpha)+\alpha^2\Gamma(\alpha)+\Gamma(1+\alpha)-\alpha\Gamma(1+\alpha)}{2\pi(-1+\alpha)\alpha\Gamma(1+\alpha)} \right).
\end{align*}
All these lead to the following relation for the perturbations outside the horizon:
\begin{equation}  \label{eq3.17}
   R_{q}(\tau) \longrightarrow R^{o}_{q}(\tau)  = \frac{-i K'}{8\pi} \Gamma(\alpha) q^{-\alpha} \left[1-\frac{K\pi}{4} \mathcal{F}_{\alpha} \right].
\end{equation}
The parameter $\alpha$ will approach to $3/2$ by applying the slow roll condition, $ \epsilon \ll 1 $.
Since $ -q \tau \ll 1 $, in the slow roll inflation, the terms $(-q\tau)^{4\alpha-3}=(-q\tau)^3 $ and $ (-q\tau)^{4\alpha-1}=(-q\tau)^5$ don't have any significant contribution in $\mathcal{F}_{\alpha}$.
‌Beside, if we consider the case in which  the non--improved solution at $\tau_0$ satisfies the initial condition  $ R_{q_{(0)}}(\tau_0) = 0$, we obtain $ \mathcal{D}_{1(\alpha)}(q) = -\mathcal{C}_{\alpha}(-q\tau_{0})^{2\alpha-1}$.
The gauge invariant quantity with this initial condition becomes
\begin{equation} \label{eq3.19}
  R^{o}_{q}(\tau) = \eta q^{-3/2} \left( 1 - \frac{K}{4\pi}\mathcal{C}_{\frac{3}{2}}\left[(-q\tau)^2-(-q\tau_0)^2\right] \right)
\end{equation}
where $\eta = -i\frac{K'}{8\pi}\Gamma(\frac{3}{2})$ is a constant with the dimension of $[L]^{-1}$ and $\mathcal{C}_{\frac{3}{2}} = -\frac{4i}{6\pi}$.
Clearly the second term in the parenthesis is the consequence of improvement.

Before continuing, we have to note that, here the IMSE \eqref{eq2.24}, for inflation with exponential potential $V(\varphi) = g e^{-\lambda\varphi}$ was solved. 
One may be wondering about how the results depends on the specific inflationary model we used. Therefore, it seems appropriate to take a look at the other models such as Starobinsky \cite{Starobinsky} or chaotic inflation \cite{Linde}. The former, which seems to have a compatible description of CMB, is the result of a specific 
$f(R)$ gravity (with $f(R) = R + R^2/6M^2$).
The Starobinsky model in the Einstein frame (obtained by the conformal transformation $ g_{\mu\nu} \rightarrow \tilde{g}_{\mu\nu} = e^{\sqrt{\frac{16\pi G}{3}}\varphi} g_{\mu\nu} $)  is described by the action
 \begin{equation}
 S = \int{d^{4}x\sqrt{-\tilde{g}}(\frac{1}{16\pi G}\tilde{R}-\dfrac{1}{2}\tilde{g}^{\mu\nu}\partial_{\mu}\varphi \partial_{\nu}\varphi-V(\varphi))}
\end{equation}
where $V(\varphi)=3M^2(1-e^{-\sqrt{\frac{16\pi G}{3}}\varphi})^2 / 32\pi G$.  For the  FRW background this leads to an inflationary scale factor $a(t) \sim e^{t^2/12 t_{pl}^2}$. Since we are interested in the effects of improvement at the \textit{horizon exit time}, and expanding $t\simeq t_0+\delta t $ (with $t_0$ much larger than $t_{pl}$), the behavior of the scale factor at times that we are interested in is given by
\begin{equation}
a(t) \sim \exp\left(\frac{t_{0}\delta t}{6t_{pl}^{2}}\right )
\end{equation}
This is a similar behavior as the one obtained by the exponential potential
\begin{equation}
a(t) \sim (t/t_{pl})^{1/\epsilon} \sim  \exp\left(\frac{\bar{t}_{0}\delta t}{\epsilon t_{pl}^{2}}\right )
\end{equation}
A similar treatment holds for the chaotic inflation models with potential  $V(\varphi) = g \varphi^n$.  For example the scale factor for typical $n=2$ is given as$a(t) \sim e^{-\alpha (\beta+ t/t_{pl})^2}$, leading to the same behavior for times that we are interested in. 

This is what is expected, since all the inflationary models are built to purge the effects of the initial conditions, although they have different scale factors at the start of the inflation era.

As a result of the above discussion, one can conclude that although the details of the improved gauge invariant quantity (\ref{eq3.19}) depends on the inflationary model used,  but the general behavior is the same. Because of this, we confine ourselves to relation (\ref{eq3.19}) in the next sections.
\subsection{Case II: $\Lambda$-inflation}
For a $\Lambda$-inflation model we consider the running gauge coupling, $\Lambda(t)$ and not the $\varphi$-field, as the agent that produces exponential expansion in the inflation era. Although this not a good model (because of giving a non--improved spectral index $n_s=-2$), but we investigate it to see how the improvement affects such a model.  

For simplicity, here we use the IMSE in terms of time $t$, instead of the cosmological time:
\begin{equation} \label{eq3.22}
  \frac{d^2R_{q}}{dt^{2}} + S(t) \frac{dR_{q}}{dt} +\frac{q^{2}}{a^{2}} R_{q} =0
\end{equation}
in which 
\begin{equation} \label{eq3.23}
    S(t) = 3 H - \frac{2 \dot{H}}{H} + \frac{\ddot{H}}{\dot{H}} + \frac{\dot{G}(t)}{2 G(t)} \frac{H}{\dot{H}} \frac{d}{d t} \ln \left(\frac{G(t)^{2} \dot{H}}{\dot{G}(t) H^{3}}\right) .
\end{equation}

If we assume that the inflation field only produce the scalar perturbations and does not affect the expansion rate of universe, we can neglect $ \rho_\varphi$ with respect to $\rho_\Lambda$.
Substituting $ H = \sqrt{\frac{\Lambda(t)}{3}}$ and $ G(t) $ and $ \Lambda(t) $ from \eqref{eq1.6} and \eqref{eq1.7} in \eqref{eq3.23}, we get: 
\begin{equation} \label{eq3.24}
   \begin{split} 
    S(t) \simeq  & \quad +\frac{a_{0} e^{\sqrt{\frac{\Lambda(t)}{3}} t}}{t}
      \Bigl[ \alpha_1\beta_2 \left(\frac{t_{pl}}{t}\right)^{-2} + \left( \alpha_{1} \gamma_{2}+\beta_{1}\beta_{2} \right) + \left( \alpha_{1} \delta_{2}+\beta_{1} \gamma_{2} + \gamma_{1}\beta_{2} \right)\left(\frac{t_{pl}}{t}\right)^{2} + {} \\  
   & \left(\alpha_{1} \Sigma_{2} + \beta_{1} \delta_{2} + \gamma_{1} \gamma_{2} + \delta_{1} \beta_{2} \right)\left(\frac{t_{pl}}{t}\right)^{4} \Bigr] + \sqrt{3\Lambda_{0}} - \frac{5}{t} + \mathcal{O}\left(\left(\frac{t}{t_{pl}}\right)^{-5}\right )
    \end{split}
\end{equation} 
where
\begin{align*} \label{eq3.25}
      & \alpha_{1} = -\frac{\Lambda_{0} \tilde{\omega}}{2 \tilde{\nu} m_{pl}^{2}} \quad, 
      \quad \beta_{1}  = -\frac{\Lambda_{0}}{2 \tilde{\nu} m_{pl}^{2}}  ( -2 \Omega + \tilde{\omega^{2}} ) \quad, 
      \quad \beta_{2} = -2 \\
      & \gamma_{1} = -\frac{\Lambda_{0}}{2 \tilde{\nu} m_{pl}^{2}} ( -3 \Omega \tilde{\omega} + \frac{\tilde{\nu}\tilde{\omega}m_{pl}^{2}}{\Lambda_0} ) \quad, 
      \quad \gamma_{2} = -4 ( \frac{\Omega}{\tilde{\omega}} - \tilde{\omega} ) \quad,
      \quad \Sigma_{2} = -80 \Omega(\frac{\Omega}{\tilde{\omega}}-\omega) \\
      & \delta_{1} =  -\frac{\Lambda_{0}}{2 \tilde{\nu} m_{pl}^{2}} ( 2 \Omega^{2} - 2 \Omega \frac{\tilde{\nu}m_{pl}^{2}}{\Lambda_{0}} +\frac{\tilde{\omega}^{2} \tilde{\nu}m_{pl}^{2}}{\Lambda_0} ) \quad,
      \quad \delta_{2} = 16(\Omega+(\frac{\Omega}{\tilde{\omega}}-\tilde{\omega})^{2}).
\end{align*}
In terms of the dimensionless time $ T =\frac{t}{t_{pl}}$ this reads as: 
 \begin{equation} \label{eq3.26}
    S(T) \simeq  \quad +\frac{a_{0}}{T} \Bigl[ z_{1} T^{2} +z_{2} + z_{3}T^{-2} +z_{4}T^{-4} \Bigr] + \sqrt{3\Lambda_{0}}t_{pl} - \frac{5}{T} + \mathcal{O}\left(\left(\frac{t}{t_{pl}}\right)^{-5}\right ).
\end{equation} 
where $z_{1} = \alpha_1\beta_2$, $z_{2} = \alpha_{1} \gamma_{2}+\beta_{1}\beta_{2}$, $z_{3} = \alpha_{1} \delta_{2}+\beta_{1} \gamma_{2} + \gamma_{1}\beta_{2}$ and $z_{4} = \alpha_{1} \Sigma_{2} + \beta_{1} \delta_{2} + \gamma_{1} \gamma_{2} + \delta_{1} \beta_{2}$.
Since at the end of inflation:
\begin{equation*}
 z_{1}T_{EI}^2\simeq 1.32\times10^{-12},\quad z_{2}\simeq -1.57\times10^{-36},\quad z_{3}T_{EI}^{-2}\simeq 1.19\times10^{-24},\quad z_{4}T_{EI}^{-4} \simeq -1.42\times10^{-48 },
\end{equation*}
we can ignore $z_{2}$, $z_{3} T^{-2}$, $z_{4} T^{-4}$ and the constant $ \sqrt{3\Lambda_{0}}t_{pl} $. As a result the IMSE in this case is given by:
\begin{equation} \label{eq3.27}
  \frac{d^{2}R_{q}}{dT^{2}} -\frac{5}{T}  \frac{dR_{q}}{dT} +\mathcal{Q}^{2}_{q} R_{q} =  a_{0} z_{1} T  \frac{dR_{q}}{dT}
\end{equation}
where 
\begin{equation}
\mathcal{Q}_{q}=\frac{qt_{pl}}{a}\simeq \frac{qt_{pl}}{a_{0}}.
\end{equation}
It has to be noted that the term $\frac{5}{T}$ comes from $a\frac{d}{dt} \ln(a^2\sqrt{\frac{3}{\Lambda}}\frac{\dot{\Lambda}}{2\Lambda})$ and has the most contribution in this model. 
In contrast, $-a_{0} z_{1} T$ is a consequence of the running gravitational coupling.

In a very similar way like the previous subsection the solution to \eqref{eq3.27} can be obtained as:
\begin{equation} \label{eq3.28}
  R_{q_{(0)}}(T) = K'' T^{3} H^{(1)}_{3} (\mathcal{Q}_{q} T)
\end{equation}
where $ K'' $ is a constant.
It is notable that the parameter $\alpha$ in Hankel function $H^{(1)}_{\alpha}(\mathcal{Q}_{q} T)$ is fixed now.

As for the previous model the solution in the next order of iteration can be obtained leading to the following perturbations outside the horizon:
\begin{equation} \label{eq3.29}
   R_{q}(T) \longrightarrow R^{o}_{q}(T)  = \frac{-i K''}{8\pi} \Gamma(3) \mathcal{Q}_{q}^{-3} \left[1+\frac{a_{0} z_{1}\pi}{4} \mathcal{F}'_{3} \right]
\end{equation}
where
\begin{equation}
  \mathcal{F}'_{3} = -\frac{i}{24\pi}\mathcal{Q}_{q}^{-2}[(\mathcal{Q}_{q}T)^{4}-(\mathcal{Q}_{q}T_{0})^{4})].
\end{equation}
Putting everything in place, the gauge invariant quantity would be:
\begin{equation} \label{eq3.29}
   R^{o}_{q}(T)  = \frac{-i K''}{4\pi} \mathcal{Q}_{q}^{-3} \left[1-i\frac{a_{0} z_{1}}{96} \mathcal{Q}_{q}^{-2}[(\mathcal{Q}_{q}T)^{4}-(\mathcal{Q}_{q}T_{0})^{4})] \right].
\end{equation}

In the next section, we shall investigate the effect of these improvements on the gauge invariant quantity on the power spectrum of perturbations for Case I which is a realistic model.
\section{Improved power spectrum}\label{Improved power spectrum}
The freezing of gauge invariant quantity after exiting from the horizon proposes cosmological observable quantities such as correlation function, $\Phi(q)$ which is the Fourier transform of 2-point function and gives us some information about the inhomogeneities.
It is given by the relation:
\begin{equation}
\langle R_q R^{*}_{q'}\rangle = (2\pi)^{3} \delta(q+q') \Phi(q).
\end{equation}
Using the results of the previous section we can easily obtain the correlation function for the first case ($\Lambda_0=0$) as
\begin{equation}
\Phi(q)\propto q^{-3}\left | \left ( 1+\frac{iK}{6\pi^2} \left(\frac{q}{q_0}\right)^{2} \left ( 1-(\frac{\tau_0}{\tau})^2 \right )\right ) \right |^{2}
\end{equation}
in which $q$ is normalized to $q_{0}=1/\tau$. The normalized correlation function both in the non--improved and improved cases are plotted in figure (\ref{fig1}). 
\begin{center}
	\includegraphics[width=0.80\textwidth]{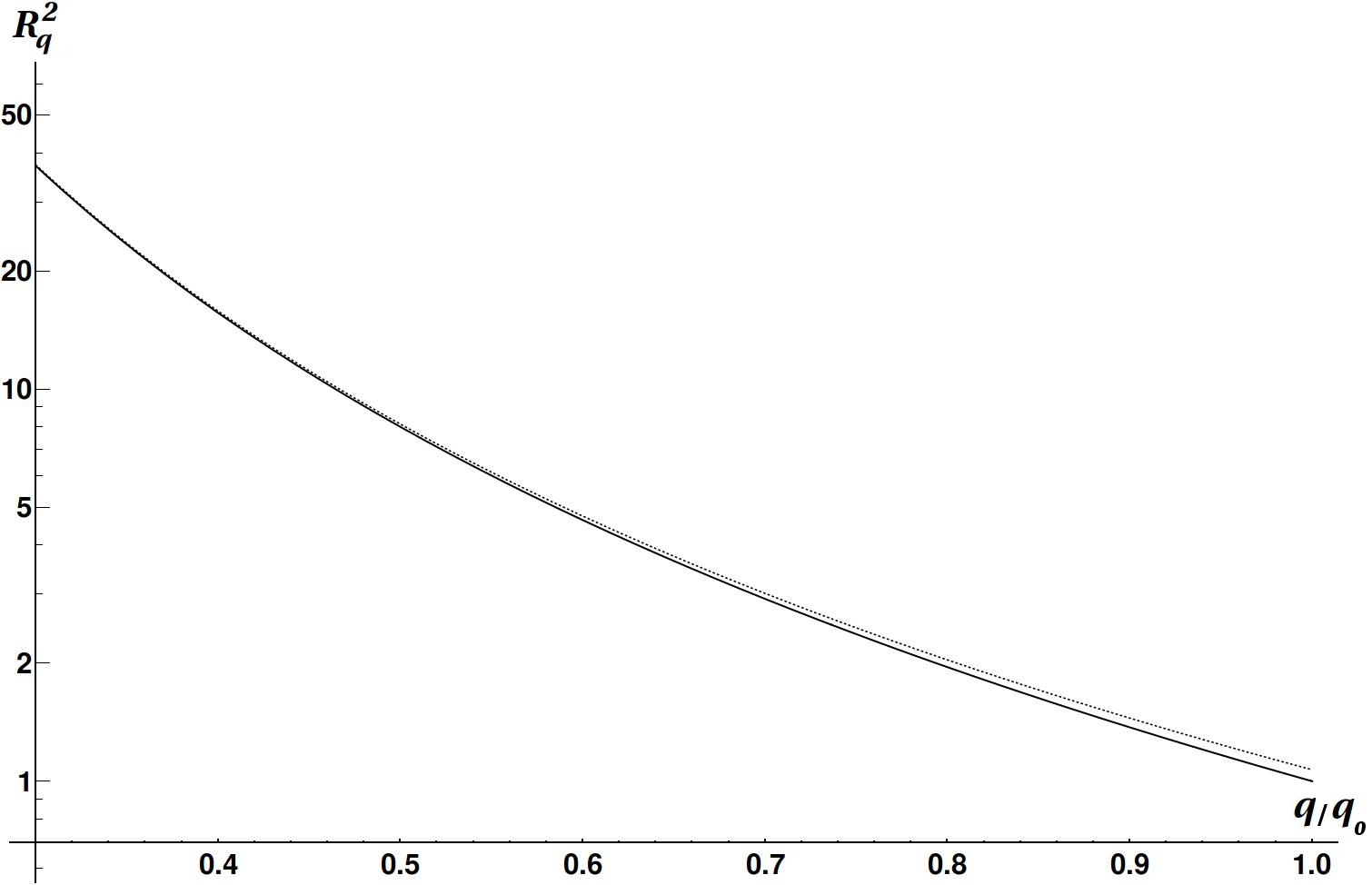}
	\captionof{figure}{The normalized non--improved (solid line) and improved (dotted line) correlation function versus the normalized wave number ($\text{q}$), with $\epsilon=0.3$.}
	\label{fig1}
\end{center}
One can see in this plot that  the effect of improvement coming from the  running couplings is to have a little larger correlation for large wave numbers.

The relative correction of the correlation function is shown in figure (\ref{fig2}) for three different values of the slow roll parameter. It shows that the correction becomes smaller as the slow roll parameter decreases. This correction is about (for $q=q_{0}$) $5$ \% for  $\epsilon=0.300$, $5\times10^{{-2}}$ \% for $\epsilon=0.030$ and $5\times10^{-4}$ \% for $\epsilon=0.003$.
\begin{center}
	\includegraphics[width=0.80\textwidth]{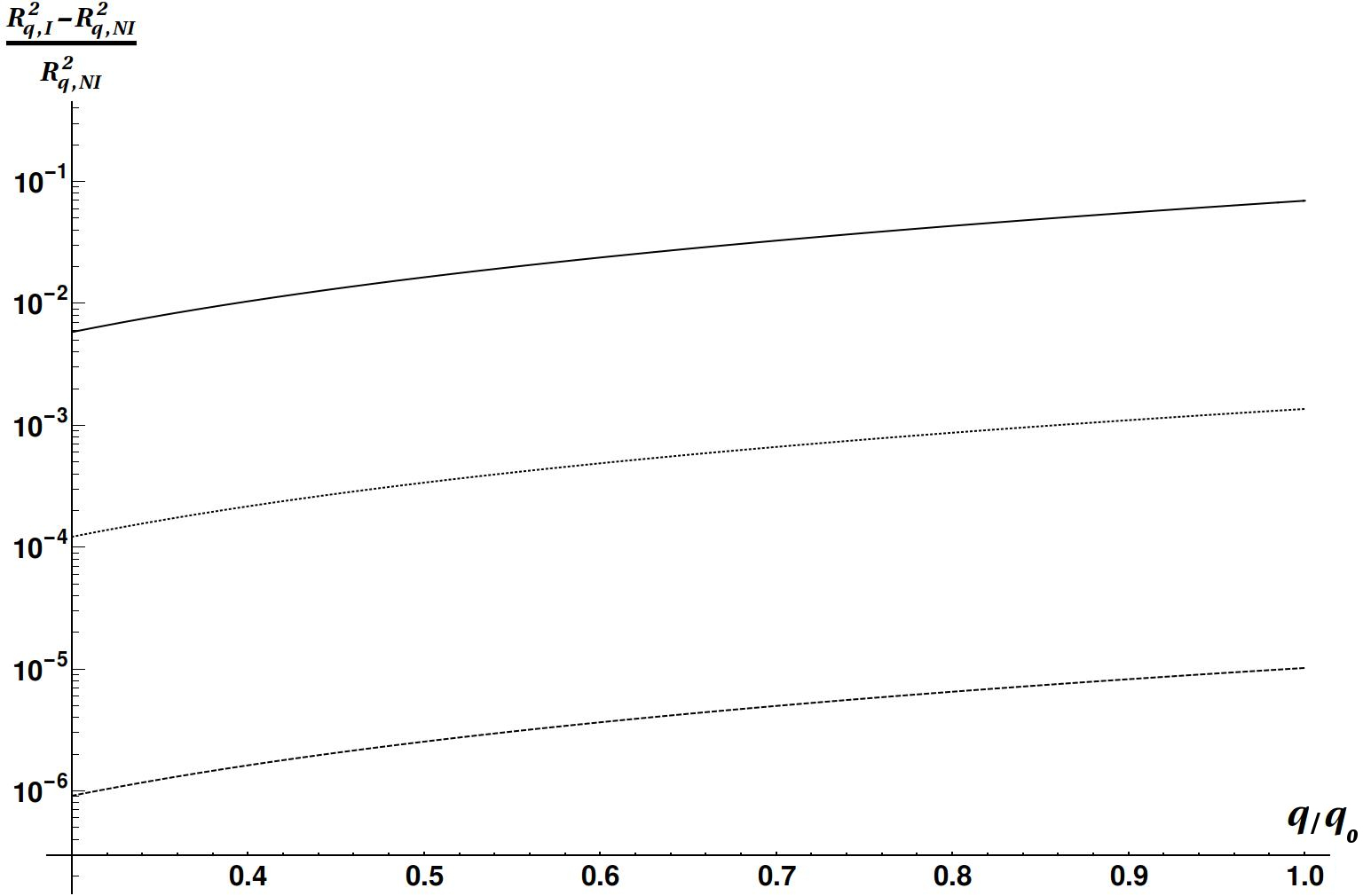}
	\captionof{figure}{Relative deviation of improved correlation function from the non--improved one versus the normalized wave number.
	$\epsilon=0.300$ for solid curve, $\epsilon=0.030$ for dotted curve, and $\epsilon=0.003$ for dashed curve.}
	\label{fig2}
\end{center}

In order to obtain the power spectrum, note that the improved form of $G(t)$ \eqref{eq1.6} quickly approaches to $G_0$ for larger times, and thus can be considered constant after inflation and in the radiation and matter dominated era. \textit{This means that the non--improved growth of perturbations works also in this case} and the transfer function has not any improvement and is given by the standard one \cite{Weinberg-Cosmology}. Therefore the power spectrum is given by:
\begin{equation}
\mathcal{P}(\kappa)\propto \kappa \mathcal{T}^{2}(\kappa) \Phi(\kappa)
\end{equation}    
where $\mathcal{T}$ is the transfer function \cite{Weinberg-Cosmology}, $\kappa=\sqrt 2 q/q_{EQ}$, $q_{EQ}\simeq \Omega_{M}h^{2}/13.6\text{Mpc}$ (with $\Omega_{M}$ the matter density parameter, and $H=h\times 100\text{km/sec Mpc}$) is the exit wave number at radiation--matter equality time, and
\begin{equation}
\Phi(\kappa) \propto  1+\frac{\textbf{Im}(K)}{6\pi^2}\kappa^2+\frac{\textbf{Re}(K)^2+\textbf{Im}(K)^2}{144\pi^4}\kappa^4.
\end{equation}
The result is plotted in figure (\ref{fig3}) and the improved power spectrum is compared with the non--improved one. 
\begin{center}
	\includegraphics[width=0.80\textwidth]{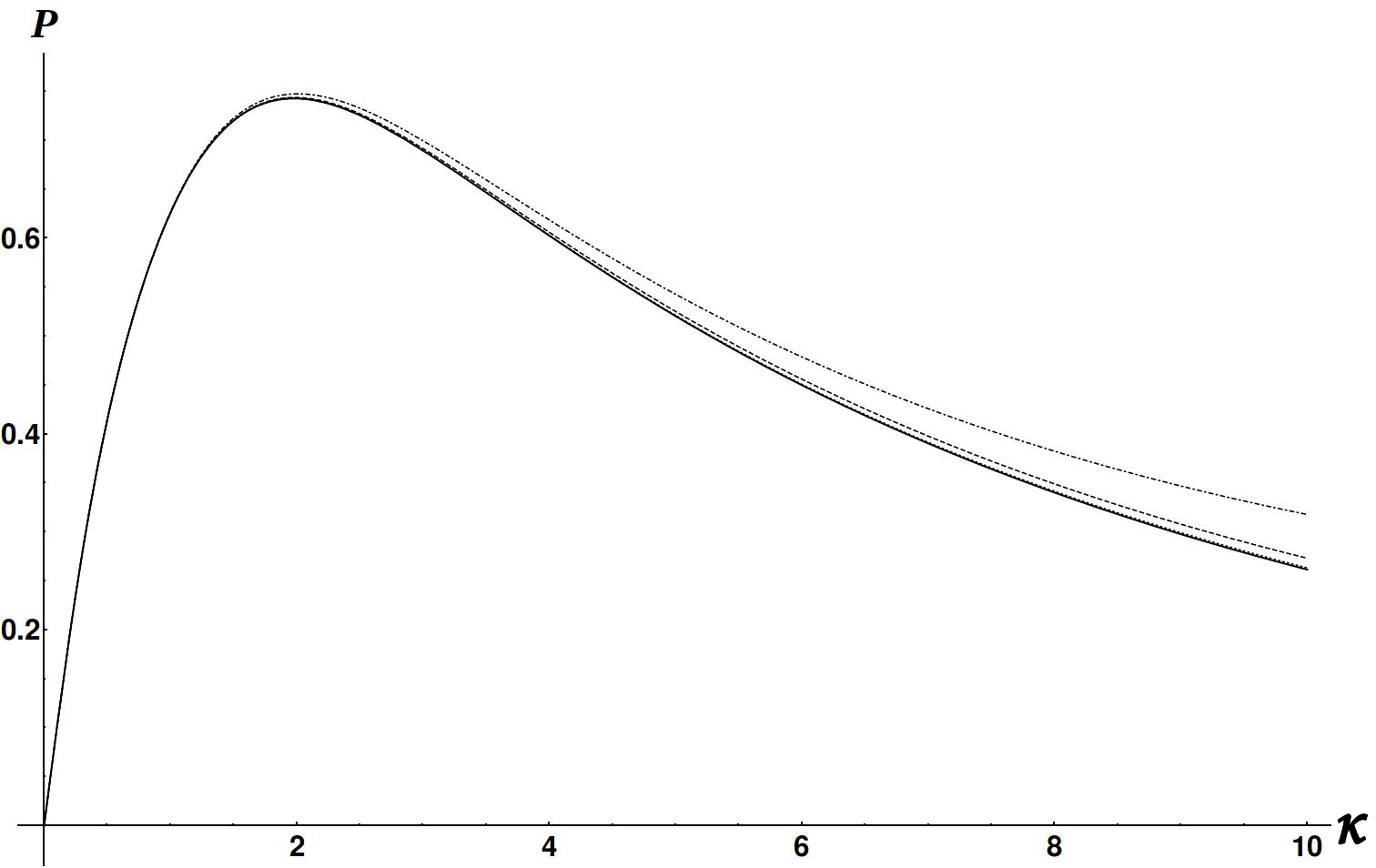}
	\captionof{figure}{Comparison of improved power spectrum with the non--improved one.
	Solid curve is the standard non--improved result, dotted curve is improved with $\epsilon=0.010$, dashed curve with $\epsilon=0.025$, and  dot--dashed curve with $\epsilon=0.050$.}
	\label{fig3}
\end{center}

As it can be seen from this plot, the RG improvement changes slightly the power spectrum for large wave numbers.  In order to see how much such an improvement is,  the observational data and the obtained results are compared in figure (\ref{fig4}).

The observational data are from \cite{Percival}, the solid curve is the non--improved power spectrum with the spectral index $n_s=0.967$ ($\kappa^{n_s}\mathcal{T}(\kappa)$). The dashed curve is the prediction of the improved model considered in this paper with the slow roll parameter $\epsilon=0.008$. In both curves we choose $\Omega_Mh=0.16$.

As it is seen from the results, the effect of improvement is to bring the tail of the graph up and leads to more accurate fit with the observed results.
\begin{center}
	\includegraphics[width=0.80\textwidth]{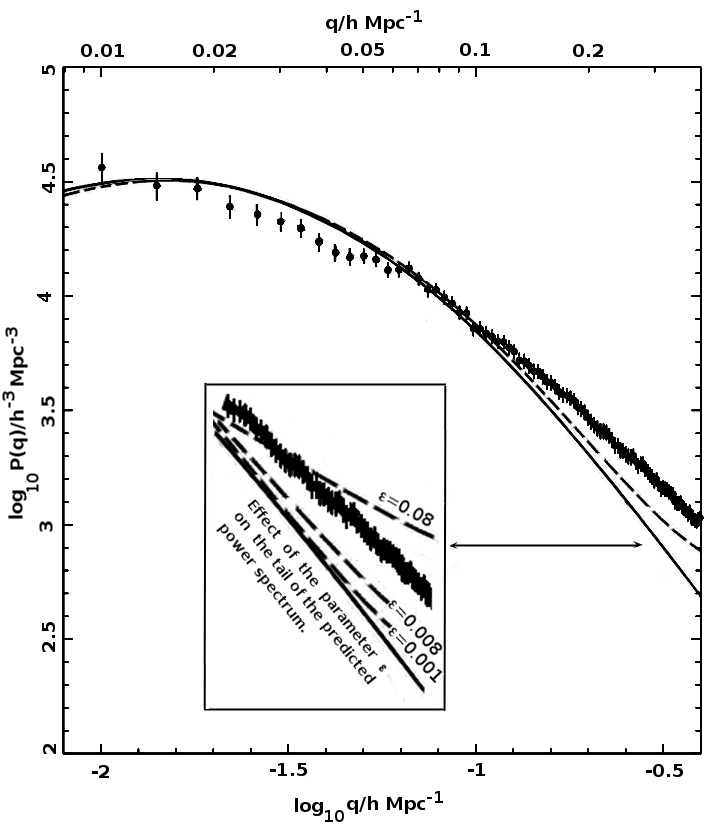}
	\captionof{figure}{Comparison of improved power spectrum with observational data.
	Solid curve is the standard non--improved result with $n_s=0.967$ and $\Omega_Mh=0.16$, dashed curve is the improved result with $\epsilon=0.008$ and $\Omega_Mh=0.16$. The main effect of increasing the parameter $\epsilon$ is to raise the tail of the improved power spectrum, as it is shown in the closeup.}
	\label{fig4}
\end{center}
\section{Conclusions}
As the asymptotically safe gravity leads to running gravitational and cosmological couplings, it is a natural question to look for its effect on the cosmological models. The effect of RG improvement of the couplings as a result of asymptotic safety becomes more important at early universe.

Here we have investigated the effect of running couplings on the curvature invariant as the seed for initial perturbations, by deriving the improved Mukhanov--Sassaki equation. We saw that this can be solved iteratively. 

The gravitational coupling is almost constant in the radiation and matter dominated eras and thus the transfer function of perturbations to the power spectrum is not improved. We obtained the improved power spectrum and observed that it is slightly improved for large wave numbers. 
This means that the tail of the predicted power spectrum comes more closer to the observed data.


\begin{thebibliography}{}
\bibitem{Kiefer} C. Kiefer, \textit{Quantum Gravity}, Oxford University Press  (2007).
\bibitem{Rovelli} C. Rovelli, \textit{Qunautm Gravity} (Cambridge Monographs on Mathematical Physics), Cambridge: Cambridge University Press (2008).
\bibitem{1st Weinberg} S. Weinberg, in \textit{Understanding the Fundamental Constituents of Matter}, edited by A. Zichichi, New York: Plenum Press (1978).\\ 
S. Weinberg, in \textit{General Relativity}, edited by S. W. Hawking and W. Isreal, Cambridge: Cambridge University Press (1979).
\bibitem{Weinberg inflation} S. Weinberg, \textit{Phys. Rev. D}, \textbf{81}, 083535 (2010).
\bibitem{1st Reuter} M. Reuter, \textit{Phys. Rev. D}, \textbf{57}, 971 (1998). \\ 
M. Reuter, arXiv: hep-th/9605030.
\bibitem{Finite Theory Space} A. Codello, R. Percacci and C. Rahmede, \textit{Int. J. Mod. Phys. A}, \textbf{23}, 143 (2008).
\bibitem{Varing G} V. Faraoni, \textit{Cosmology in Scalar--Tensor Gravity} (Fundamental theories of Physics), Springer (2004).
\bibitem{IBH} M. Reuter and E. Tuiran, \textit{Phys. Rev. D}, \textbf{83}, 044041 (2011).\\
D. Becker and M. Reuter, \textit{JHEP}, \textbf{1207}, 172 (2012).
\bibitem{IGR} M. Reuter and H. Weyer, \textit{Phys. Rev. D}, \textbf{70}, 124028 (2004).
\bibitem{Agullo}I. Agullo, J. Navarro--Salas, G. J. Olmo, and L. Parker \textit{Phys. Rev. Lett}, \textbf{103}, 061301 (2009), \\
I. Agullo, J. Navarro--Salas, G. J. Olmo, and L. Parker \textit{Phys. Rev. D}, \textbf{81}, 043514 (2010).
\bibitem{Bonanno & Reuter1} A. Bonanno and M. Reuter, \textit{Phys. Rev. D}, \textbf{62}, 043008 (2000).
\bibitem{Nagy} S. Nagy, \textit{Ann. Phys}, \textbf{350}, 310 (2014).
\bibitem{Bonanno & Reuter2} A. Bonanno and M. Reuter, \textit{Phys. Rev. D}, \textbf{65}, 043508 (2002).
\bibitem{Bonanno11} A. Bonanno, \textit{Class. Quant. Grav.}, \textbf{28}, 145026 (2011), \\
A. Bonanno, \textit{PoS CLAQG}, \textbf{08}, 008 (2011).
\bibitem{Falls} K. Falls, F. Litim and A. Raghuraman, \textit{Int. J. Mod. Phys. A}, \textbf{27}, 1250019 (2012).
\bibitem{Ellis} G. F. R. Ellis and J. Uzan, \textit{Am. J. Phys.}, \textbf{73}, 240 (2005).
\bibitem{Reuter & Weyer} M. Reuter and H. Weyer, \textit{Phys. Rev. D}, \textbf{69}, 104022 (2004).
\bibitem{Weinberg-Cosmology} S. Weinberg,  \textit{Cosmology}, Oxford University Press (2008).
\bibitem{Percival}W. J. Percival, \textit{et al}, \textit{Astrophys. J.}, \textbf{657}, 645 (2007).
\bibitem{Starobinsky} A. A. Starobinsky, \textit{Phys. Lett. B}, \textbf{91}, 99–102 (1980).
\bibitem{Linde} A. D. Linde, \textit{Phys. Lett. B}, \textbf{129}, 177-181, (1983).
\end{thebibliography}
\end{document}